\documentclass[aps,prb,preprint,superscriptaddress,epsfig,showpacs]{revtex4}
\usepackage{graphicx}
\usepackage{amsmath}

\begin{document}

\title{Effect of the Coulomb interaction on the electron relaxation
of weakly-confined quantum dot systems}

\author{Juan I. Climente}
\affiliation{CNR-INFM S3, Via Campi 213/A, 41100 Modena, Italy}
\email{climente@unimo.it}
\homepage{http://www.nanoscience.unimo.it/}
\author{Andrea Bertoni}
\affiliation{CNR-INFM S3, Via Campi 213/A, 41100 Modena, Italy}
\author{Massimo Rontani}
\affiliation{CNR-INFM S3, Via Campi 213/A, 41100 Modena, Italy}
\author{Guido Goldoni}
\affiliation{CNR-INFM S3, Via Campi 213/A, 41100 Modena, Italy}
\affiliation{Dipartamento di Fisica, Universit\`a degli Studi di Modena e Reggio Emilia,
Via Campi 213/A, 41100 Modena, Italy}
\author{Elisa Molinari}
\affiliation{CNR-INFM S3, Via Campi 213/A, 41100 Modena, Italy}
\affiliation{Dipartamento di Fisica, Universit\`a degli Studi di Modena e Reggio Emilia,
Via Campi 213/A, 41100 Modena, Italy}
\date{\today}

\begin{abstract}
We study acoustic-phonon-induced relaxation of charge excitations 
in single and tunnel-coupled quantum dots containing few confined 
interacting electrons.
The Full Configuration Interaction approach is used to account 
for the electron-electron repulsion.
Electron-phonon interaction is accounted for through both
deformation potential and piezoelectric field mechanisms. We show
that electronic correlations generally reduce intradot and interdot
transition rates with respect to corresponding single-electron
transitions, but this effect is lessened by external magnetic fields.
On the other hand, piezoelectric field scattering is found to become
the dominant relaxation mechanism as the number of confined
electrons increases. Previous proposals to strongly suppress
electron-phonon coupling in properly designed single-electron
quantum dots are shown to hold also in multi-electron devices.
Our results indicate that few-electron orbital degrees of
freedom are more stable than single-electron ones.

\end{abstract}


\maketitle
\section{Introduction}

Recent experiments have appointed coupling to acoustic phonons as
the main source of electron relaxation in the excited states of
semiconductor quantum dots (QDs) with weak lateral
confinement\cite{FujisawaNAT,FujisawaJPCM} and coupled quantum dots
(CQDs) with small tunneling
energy,\cite{FujisawaSCI,TaruchaME,OrtnerPRB} i.e., with the
electronic relevant energy scale in the few-meV range. This usually
restricts lifetimes to the order of nanoseconds,\cite{FujisawaNAT}
imposing severe limitations on the performance of QD devices which
rely on the coherent dynamics of electron charge states. A prominent
example are charge qubits, whose main disadvantage, as compared to
spin qubits, is the faster decoherence rate.\cite{FujisawaNAT}

As a first step to eventually gain control over charge relaxation
rates, several theoretical works have investigated the fundamental
physics of electron-phonon coupling in QD structures. Bockelmann
described the qualitative effect of lateral (spatial and magnetic)
confinement on the electron transition rates in
QDs.\cite{BockelmannPRB} The domains of deformation potential (DP)
and piezoelectric field (PZ) interactions, the two main scattering
mechanisms leading to interaction with acoustic phonons, were
established for both QD and CQD
structures.\cite{ClimentePRB,StavrouPRB,WuPRB,Stano_arxiv} Building on these
works, methods to minimize electron-phonon coupling in QDs and
vertically CQDs were recently proposed which may bring electron
lifetimes in the range of microseconds, thus making them comparable
to usual spin relaxation
rates.\cite{ClimentePRB,ZanardiPRL,BertoniAPL, BertoniPE}

Noteworthy, all of the aforementioned experimental and theoretical
works studied charge relaxation at a single-electron (SE) level. So
far, multi-electron (ME) systems have been essentially restricted to
the context of spin
relaxation,\cite{FujisawaNAT,FujisawaJPCM,SasakiPRL,TaylorArXiv} and
only lately charge relaxation processes in ME systems started to be
considered.\cite{BraskenMP,BertoniPRL} In particular, in
Ref.~\onlinecite{BertoniPRL} we have investigated Coulomb correlated
systems, and we have reported numerical evidence that electronic
correlations generally lead to reduced decay rates of the excited
electronic states in weakly confined dots, as compared to
independent-particle estimates. This finding
suggests that ME devices might better exploit the discrete energy
spectrum of QDs.

In this paper, we extend our previous investigation
(Ref.~\onlinecite{BertoniPRL}) analyzing in detail the mechanisms by
which Coulomb interaction affects charge relaxation rates in ME QD
systems. We also investigate the effect of external magnetic fields
on intradot transition rates, and isospin transition rates in ME
vertically CQDs. Electron-phonon interaction is included
through both DP and PZ scattering channels (the latter was neglected
in Ref. \onlinecite{BertoniPRL} as well as in many theoretical
investigations).


The paper is organized as follows. In Section \ref{sec:theo} we
describe our theoretical model. In Section \ref{sec:sqd} we study ME
charge relaxation in single QDs as a function of the lateral
(spatial and magnetic) confinement, and interpret the
correlation-induced scattering reduction in terms of the SE
configuration mixing. The applicability in ME QDs of
magnetic-field-based suggestions for controlling charge relaxation
rates, previously reported for SE
structures\cite{ClimentePRB,BertoniAPL,BertoniPE}, is addressed in
this section. In Section \ref{sec:cqd} we investigate the effect of
interdot distance on the isospin transition rate of ME vertically
CQDs, and finally in Section \ref{sec:summ} we summarize our
results.

\section{Theoretical formulation}
\label{sec:theo}

The theoretical model we use is similar to that described in
Ref.~\onlinecite{ClimentePRB} for single-electron excitations, but
now considering $N$-electron states ($N=1-5$).
 We study disk-shaped QDs where the lateral confinement is much weaker than
the vertical one, and the dot and surrounding barrier are made of
materials with similar elastic properties. 
Several QD structures reported in the literature fit this
description.\cite{FujisawaNAT,FujisawaJPCM,weakdots}

A number of useful approximations can be made for such QDs. First,
since the weak lateral confinement gives inter-level spacings within
the range of few meV, only acoustic phonons have significant
interaction with free carriers, while optical phonons can be safely
neglected. Second, the elastically homogeneous materials are not
expected to induce phonon confinement, which allow us to consider
only bulk phonons. Finally, the different energy scales of vertical
and lateral electronic confinement allow us to decouple vertical and
lateral motion in the building of SE spin-orbitals. We then take
a parabolic confinement profile in the in-plane $(x,y)$ direction,
with single-particle energy gaps $\hbar\omega_0$, which yields the
Fock-Darwin states.\cite{ReimannRMP} In the vertical direction ($z$)
the confinement is provided by a rectangular quantum well of length
$L_z$, with the barrier height determined by the band-offset between
the QD and barrier materials. The quantum well solutions are derived
numerically. Spin-orbit coupling is neglected in this work, since
the long spin relaxation times measured in QD structures similar to
those we study here indicate that the spin degrees of freedom are
well separated from the orbital ones.\cite{FujisawaNAT} Therefore,
the SE spin-orbitals can be written as:

\begin{equation}
\label{eq:0}
\psi_{\alpha} (x,y,z;\sigma) = \phi_{n,m}(x,y) \,\xi_g(z) \,\chi_{\sigma},
\end{equation}

\noindent where $\phi_{n,m}$ is the $n-$th Fock-Darwin orbital with
azimuthal angular momentum $m$, $\xi_g$ is the symmetric quantum
well solution with parity $g$ with respect to the reflection about
the $z=0$ plane, with $g=0\,(1)$ denoting even (odd) parity, and
$\chi_{\sigma}$ is the spinor eigenvector of the spin $z-$component
with eigenvalue $\sigma$ ($\sigma=\pm 1/2$). We will also label
Fock-Darwin states with the standard notation $n\,l$, where
$l=s,\,p_{\pm},\,d_{\pm} \ldots$ correspond to $m=0,\, \pm 1, \pm 2
\ldots$, respectively.

As for the inclusion of Coulomb interactions, we need to go beyond
the Hartree-Fock approximation in order to include electronic
correlations, which are critical in phonon-induced electron
scattering processes.\cite{BraskenMP} Moreover, since we are
interested in the relaxation time of excited states, we need to know
both ground and excited states with comparable accuracy. Our method
of choice is the Full Configuration Interaction approach: the ME
wave functions are written as linear combinations $|\Psi_a \rangle =
\sum_i c_{a i} |\Phi_i\rangle$, where the Slater determinants
 $|\Phi_i \rangle=\Pi_{\alpha_i} c_{\alpha_i}^{\dagger} |0 \rangle$ are obtained
by filling in the SE spin-orbitals $\alpha$ with the $N$ electrons
in \emph{all} possible ways consistent with symmetry requirements;
here $c_{\alpha}^{\dagger}$ creates an electron in the level
$\alpha$. The ME ground and excited states can then be labeled by
the total angular momentum $z$ component $M=0,\pm 1,\pm 2\ldots$,
total parity $G=0,1$, total spin $S$, and total spin $z$-projection
$S_z$.
The fully interacting Hamiltonian is numerically diagonalized, exploiting orbital and spin
symmetries.\cite{RontaniJPC}

We assume zero temperature, which suffices to capture the main features of one-phonon
processes.\cite{BertoniAPL} At this temperature, only phonon emission processes are present.
We evaluate the relaxation rate between selected initial (occupied) and final (unoccupied) ME states,
 $b$ and $a$, using the Fermi golden rule:

\begin{equation}
\tau^{-1}_{b \rightarrow a}=\frac{2\pi}{\hbar}\,\sum_{\nu \mathbf{q}}
\Bigl\lvert \sum_{ij} c_{bi}^* c_{aj} \langle \Phi_i|V_{\nu \mathbf{q}}|\Phi_j\rangle \Bigr\rvert^2\,
\delta(E_b-E_a - E_q),
\label{eq1}
\end{equation}

\noindent where the electron states $| \Psi_\alpha \rangle$  ($\alpha=a,b$) have been written explicitly
as linear combinations of Slater determinants, $V_{\nu \mathbf{q}}$ is the interaction
operator of an electron with an acoustic phonon of momentum $\mathbf{q}$ via deformation
potential $(\nu=DP)$ or piezoelectric field $(\nu=PZ)$ interaction, $E_\alpha$ stands for
the $\alpha$ electron state energy and $E_q$ represents the phonon energy.

The electron-phonon interaction matrix element can be written more
explicitely as

\begin{equation}
\langle \Phi_i|V_{\nu \mathbf{q}}|\Phi_j\rangle =
 M_{\nu}(\mathbf{q})\, \langle \Phi_i|\,e^{-i \mathbf{q r}}\,|\Phi_j \rangle,
\label{eq2}
\end{equation}

\noindent where the right-most term is the electron form factor and
$M_{\nu}(\mathbf{q})$ is a prefactor which depends on the scattering
mechanism $\nu$.\cite{ClimentePRB} It is worth noting that, whereas
for DP scattering $M_{DP} \propto \sqrt{|\mathbf{q}|}$, for PZ
scattering $M_{PZ} \propto 1/\sqrt{|\mathbf{q}|}$. As a result, DP
scattering is dominant when the emitted phonon energy is
sufficiently large, while PZ scattering dominates at small phonon energy.
For SE transitions in GaAs QDs, even in the weakly-confined regime, the
DP mechanism usually prevails in the absence of external fields.
However, we have recently shown \cite{ClimentePRB} that the PZ
mechanism may rapidly become dominant in the presence of a
vertical magnetic field which tends to suppress the single-particle
gaps.

One can see from the above expressions that Coulomb interaction
influences electron scattering with phonons in two ways. First, it
introduces changes in the electron energy gaps $E_b-E_a$, and hence
in the energy and momentum of the emitted phonon.\cite{BraskenMP}
Second, it introduces changes in the orbital part of the electron
state, hence changing the electron-phonon wave functions coupling.
The latter effect is reflected in Equation (\ref{eq1}) through the
Slater determinant coefficients. Indeed, since the total scattering
rate is but a weighted sum of SE contributions, the general behavior
of SE scattering events will also apply to the ME case. However, the
weight of each SE contribution strongly depends on the number of
particles, the regime of correlations, and the presence of external
fields, so that important changes in the ME relaxation rates should
be expected when varying these parameters.

In this work we consider mostly relaxation rates corresponding to
the fundamental spin-conserving transition in single and coupled
QDs, i.e. transitions involving the ground state and the first
excited state with the same $(S,\,S_z)$ quantum numbers. This
transition could be monitored, e.g., by means of pump-and-probe
techniques,\cite{FujisawaNAT,FujisawaJPCM,SasakiPRL} since
relaxation to or from any intermediate state with different spin
should be much slower and therefore it will barely interfere.

Below we shall investigate GaAs/Al$_{0.3}$Ga$_{0.7}$As QDs, using
the following material parameters:\cite{Tin_book} electron effective
mass $m^*=0.067$, band-offset $V_c=243$ meV, crystal density
$d=5310$ kg/m$^3$, acoustic deformation potential constant $D=8.6$ eV, 
effective dielectric constant $\epsilon=12.9$, and piezoelectric 
constant $h_{14}=1.41\cdot 10^9$ V/m.
For the sound speed
$c_{\sigma}$, we take into account that in cylindrical QDs most of
the scattering arises from phonon propagation close to the growth
direction.\cite{BertoniPE} We then assume that the QDs are grown
along the $[1\, 0\, 0]$ direction and use the corresponding values
$c_{\mbox{\tiny LA}}=4.72 \cdot 10^3$ m/s and $c_{\mbox{\tiny
TA}}=3.34 \cdot 10^3$ m/s.\cite{Landolt_book} In our calculations we
deal with QDs with lateral confinement energies which in some cases
are rather weak ($\hbar \omega_0 < 1$ meV). Well-converged few-body
states are obtained for such structures using a basis set composed
by the Slater determinants which result from all possible
combinations of 62 SE spin-orbitals with $N$ electrons. Due to the
strong confinement in the vertical direction, only the lowest $g=0$
(for single QDs) or the lowest $g=0,1$ (for CQDs) eigenstates are
included in the single-particle basis.

\section{Single quantum dots and magnetic field}
\label{sec:sqd}

\begin{figure}[p]
\includegraphics[width=8cm,clip]{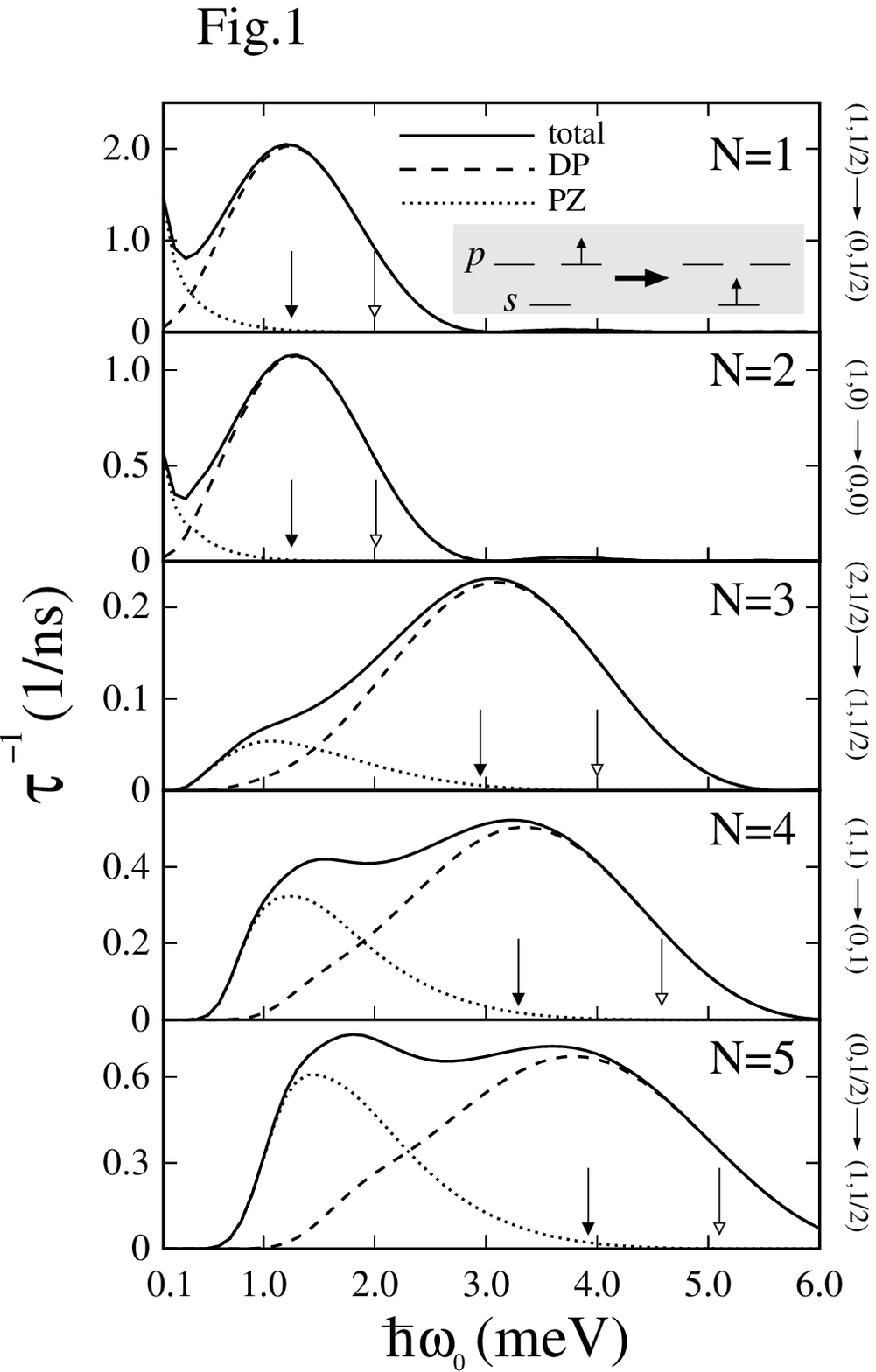}
\caption{Charge relaxation rate vs lateral confinement for a QD with
vertical width $L_z=10$ nm filled with $N=1$ to $N=5$ electrons.
Solid lines: total scattering rate. Dashed line: DP contribution.
Dotted lines: PZ contribution. Note the different vertical scale in
each panel. Next to the right axis, we indicate the quantum numbers
$(M,S)$ of the states involved in the transition. The downward
arrows in each panel point at the confinement energy leading to
emitted phonon energies of $1.3$ (solid arrowheads) and $2$ meV
(empty arrowheads). Inset in the top panel: SE configurations
involved in the $N=1$ transition. }\label{Fig1}
\end{figure}

In this section we first study charge relaxation rates in QDs with
$N$ interacting electrons as a function of the harmonic lateral
confinement originated by electrostatic fields, and next consider
the effect of adding a magnetic field. To study the lateral
confinement, we vary the characteristic frequency of the confining
parabola in the $\hbar \omega_0 \sim 0-6$ meV range, thus moving
from a strongly- to a rather weakly-correlated regime. Figure
\ref{Fig1} depicts the corresponding results for $N=1 - 5$
electrons in QDs with height $L_z=10$ nm.
It can be observed that for all $N$ the qualitative shape of the
relaxation rate curve is similar to that of the SE case (top panel):
it shows two maxima, connected with the PZ and DP scattering
mechanisms, and it vanishes at small and large confinement energies
due to the small phonon density and small electron-phonon coupling,
respectively.\cite{BockelmannPRB,ClimentePRB,noseveN1} Another trend
observed in Fig.~\ref{Fig1} is the shift of the scattering rate
maxima towards larger confinement energies with increasing number of
electrons, as well as the increasing relative height of the PZ
maximum. Both features follow from the increasing density of states
with larger $N$, which leads to smaller inter-level spacing and,
therefore, to larger $\hbar \omega_0$ values associated with a given
phonon energy. We illustrate this in Fig.~\ref{Fig1} with downward
arrows to point at the confinement energies which give two selected
phonon energies, $E_q=1.3$ (solid arrowhead) and $E_q=2.0$ meV
(empty arrowhead). It can be seen that there is a shift towards
larger $\hbar \omega_0$ values with increasing $N$.\cite{N2aN3} An
important implication of this result is that PZ scattering, which is
negligible except at very weak confinement energies in the SE
picture, may actually become the dominant scattering mechanism for
usual QD confinement energies in the ME case (see, e.g., $\hbar
\omega_0=2$ meV in the $N=5$ picture). In light of this, some of the
estimates in previous investigations, which studied electron-phonon
coupling in weakly-confined QDs considering DP interaction only, may
need a revision.\cite{BraskenMP,BertoniPRL}

One also observes in Fig.~\ref{Fig1} that the phonon energy which
gives maximum scattering is approximately constant, regardless of
the number of electrons and lateral confinement. For example, $E_q
\approx 1.3$ meV for the DP maximum. This indicates that the
scattering is mainly determined by the electron-phonon
coupling along the vertical direction: 
indeed, at $E_q \approx 1.3$ meV the longitudinal acoustic phonon
wavelength gives maximum coupling with the electron wave function
\emph{in the quantum well}, which does not depend on $\hbar
\omega_0$ and is weakly affected by Coulomb interactions.

\begin{figure}[p]
\includegraphics[width=8cm,clip]{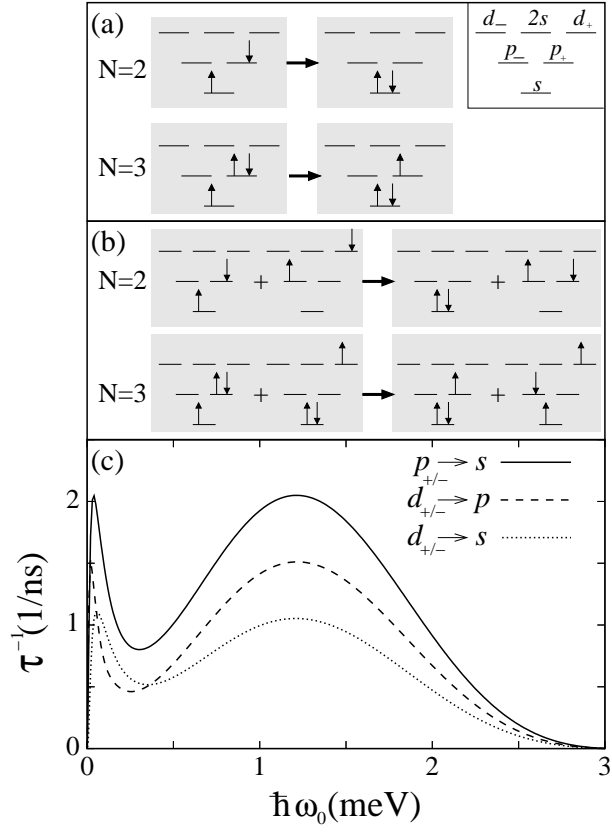}
\caption{SE electronic configurations (Slater determinants) involved
in the fundamental transition for the QDs of Fig.~\ref{Fig1} for
$N=2$ and $N=3$ for (a) non-interacting and (b) interacting
electrons. In the interacting case only the two most weighted
configurations are shown. (c) SE scattering rates vs lateral
confinement energy between selected Fock-Darwin orbitals at fixed
phonon energy $E_q=\hbar \omega_0$.}\label{Fig2}
\end{figure}

Figure \ref{Fig1} shows that the excited state lifetimes depend
strongly on the number of carriers (note the different vertical
scale of each panel), the shorter lifetimes being shown by the $N=1$
case. As a matter of fact, the $N=1$ transition, which corresponds to the $p
\rightarrow s$ relaxation, represents the independent-particle limit
of the ME cases shown in the same figure. For example, for $N=2$ an
independent particle filling gives a $(M=0,S=0)$ ground state, with
the two electrons occupying the $s$ spin-orbitals, and a $(M=1,S=0)$
excited state, with one electron in the $s$ orbital and another in
the $p$ orbital (see electronic configuration diagrams in
Fig.~\ref{Fig2}(a)). Thus, the transition involves a one-electron
scattering $p\rightarrow s$ orbital, while the other electron
remains as a spectator. Similar reasonings apply to $N=3 - 5$.
Therefore, Fig.~\ref{Fig1} shows that the Coulomb-interaction-free
relaxation rate ($N=1$ panel) generally gives an upper-bound to the
actual rate when electron-electron interaction is taken into
account. This result holds for all number of particles studied and
most confinement strengths, although the trend is non-monotonic with
$N$.

The reduction of the relaxation rate noted above can be explained in
terms of SE configurations mixing. In order to illustrate this, we
analyze in detail the $N=2$ and $N=3$ cases in Fig.~\ref{Fig2},
where panel (a) represents the electronic configurations of the
first excited and ground states which follow from an
independent-particle filling, while panel (b) represents the two
most important configurations contributing to the same states when
Coulomb interaction is included. By comparison, one can see that in
the non-interacting picture only the $p \rightarrow s$ transition
takes place, whereas in the interacting picture it is partially
replaced by the $d \rightarrow s$ and $d \rightarrow p$ transitions.
We then compare the relaxation rates of the individual SE scattering
processes [panel (c)], fixing the transition energy $E_q=\hbar
\omega_0$ in all cases in order to ensure that the comparison
considers orbital effects only. One can see that $p
\rightarrow s$ is the fastest transition. Therefore, when it is
partially replaced by $d \rightarrow s$ and $d \rightarrow p$  the
overall relaxation rate is reduced. Obviously, the stronger the
mixing of configurations the larger the reduction. This explains why
the $N=3$ scattering rate is well below that of $N=2$: the ground
state electrons already occupy the $p$-shell and the kinetic energy
difference with respect to the excited states is then much smaller,
which allows stronger Coulomb-induced mixing.
Analogous logic can be used to explain the scattering reduction in the $N>3$ systems.\\

\begin{figure}[p]
\includegraphics[width=8cm,clip]{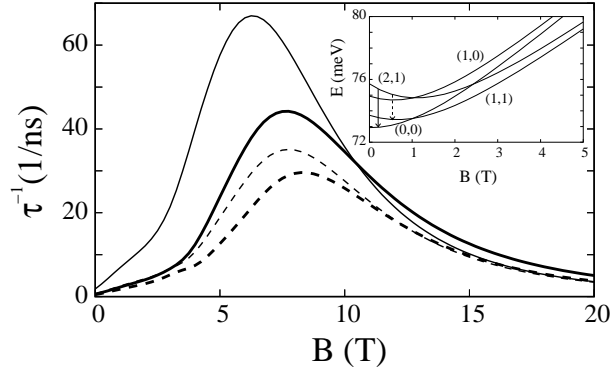}
\caption{Charge relaxation rate vs magnetic field for the lowest
singlet (solid lines) and triplet (dashed lines) transitions for
$N=2$ in a QD with $L_z=10$ nm and $\hbar \omega_0=2$ meV. Thick
lines: interacting case. Thin lines: non-interacting case. Inset:
lowest-lying energy levels and their quantum numbers $(M,S)$; arrows
indicate the transitions under study.}\label{Fig3}
\end{figure}

We next investigate the effect of a magnetic field $B$, applied
along the vertical direction of the QD, on the ME relaxation rate.
The magnetic field is expected to introduce new physics because it
strongly modifies the SE energy levels, which now draw the
well-known Fock-Darwin spectrum.\cite{ReimannRMP} In particular, the
states involved in the fundamental transition converge to the same
(lowest) Landau level. This has important implications on the energy
of the emitted phonon -which is reduced-, the regime of electronic
correlations -which become stronger- and the SE configurations of
the low-lying ME states -which differ from those at zero magnetic
field-. In Figure \ref{Fig3} we plot the relaxation rate
corresponding to $N=2$ in a QD with lateral confinement $\hbar
\omega_0=2$ meV and width $L_z=10$ nm. We show the fundamental
transition both in the singlet ($S=0$) sector (solid lines) and in
the triplet ($S=1$) sector. We also compare the interacting and
non-interacting case (thick and thin lines, respectively).

\begin{figure}[p]
\includegraphics[width=8cm,clip]{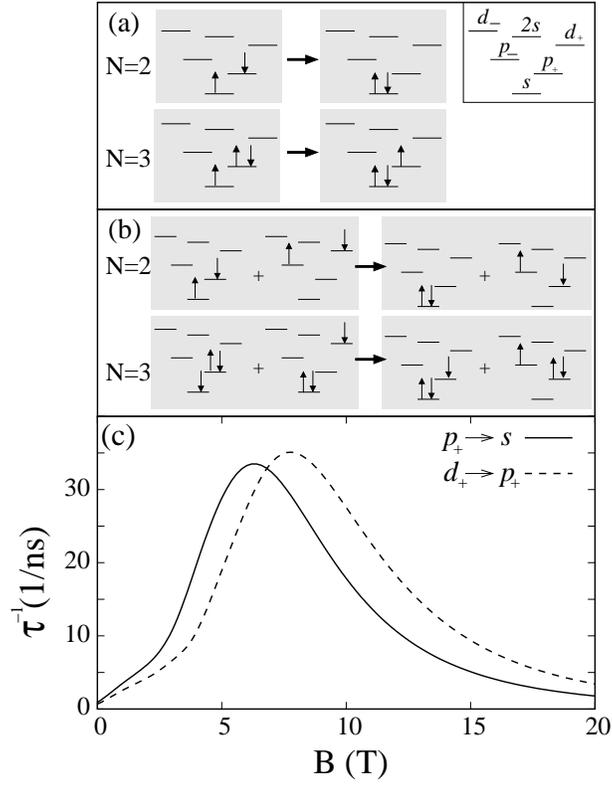}
\caption{SE electronic configurations (Slater determinants) involved
in the fundamental transition for $N=2$ and $N=3$ for (a)
non-interacting and (b) interacting electrons in a magnetic field.
In the interacting case only the two most weighted configurations
are shown. (c) SE scattering rates between selected Fock-Darwin
orbitals vs magnetic field. QD parameters are the same as in
Fig.~\ref{Fig3}.}\label{Fig4}
\end{figure}

It can be seen in the figure that the shape of the ME
curves is again qualitatively similar to that of the SE ones. From
previous investigations, we know that the scattering rate at fields
exceeding a few Tesla is largely determined by the PZ scattering
channel.\cite{ClimentePRB} In addition, we note that the
correlation-induced reduction of the relaxation rates changes with
the field. Indeed, the factor of reduction tends to decrease as
$B$ increases, and at some point ($B\approx 10.5$ T for the singlet
sector, $B \approx 13$ T for the triplet sector) the effect of
correlations is reversed, so that a small enhancement is found
instead of the more common reduction. This reversal in the behavior
can be explained in terms of the mixing of configurations in the
presence of a magnetic field. When no Coulomb interaction is
considered only the $p_+ \rightarrow s$ SE transition contributes to
the total scattering, as shown in Fig.~\ref{Fig4}(a). However, if we
take into account the Coulomb-induced mixing between the two most important
configurations, Fig.~\ref{Fig4}(b), new relaxation channels are
opened, notably the $d_+ \rightarrow s$ and $d_+ \rightarrow p_+$ SE
transitions. The increasing relative weight of these transitions
with $B$, along with the fact that the $d_+ \rightarrow p_+$
transition becomes faster than the $p_+ \rightarrow s$ one at
sufficiently strong fields [Fig.~\ref{Fig4}(c)], justify the
observation that electronic correlations at high magnetic field
yield slightly increased relaxation rates.\cite{Ecte} Yet, it may be
worth pointing out that this takes place when the states involved in
the transition have almost converged in energy (see inset in Figure
\ref{Fig3}). Indeed, at such magnetic field values the investigated
transitions are not the fundamental transition anymore, since
higher-angular-momenta states have already come down in energy (this
occurs at about $B=4$ T in the QD we study, when the $(M=2,S=1)$
level becomes the ground state).

\begin{figure}[p]
\includegraphics[width=8cm,clip]{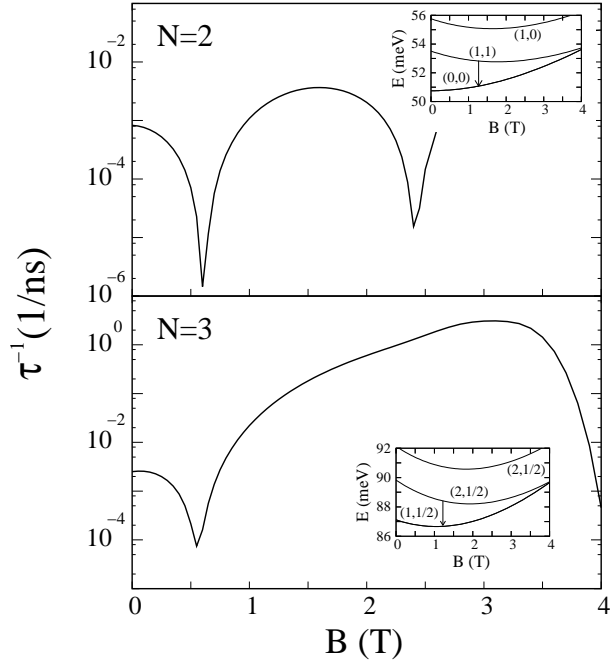}
\caption{Charge relaxation rate vs magnetic field for $N=2$ and
$N=3$ in a QD with $L_z=15$ nm and $\hbar \omega_0=5$ meV. Insets:
lowest-lying energy levels along with their quantum numbers $(M,S)$;
arrows indicate the transition under study.}\label{Fig5}
\end{figure}

Recently, it has been suggested that electron-phonon coupling can be
tailored in weakly-confined QDs and CQDs, so that SE lifetimes may be
increased by orders of magnitude.\cite{ClimentePRB,ZanardiPRL,BertoniAPL,BertoniPE}
The physical idea behind this possibility is to achieve an anti-phase relation
between the phonon wave and the electron wave function 
\emph{along the growth direction} of the QD structure,
i.e., to make the phonon wavelength along $z$ 
 a divisor of the quantum well width, thereby reducing strongly
the value of the form factor in Equation \ref{eq2}. The phonon
wavelength can be controlled through the \emph{lateral}
confinement, which is in turn determined either by electrostatic fields 
or by an axial magnetic field. 
The latter reduces the energy splitting between the low-lying SE
states and thus increases the phonon wavelength in a controllable
manner.\cite{ClimentePRB,BertoniAPL,BertoniPE} The possibility of
using external fields to suppress charge relaxation also in ME
systems may pose a significant support for the eventual fabrication
of ME QD devices with small decoherece rates. However, the
applicability is not straightforward: indeed, as mentioned before,
the energy of the emitted phonon corresponding to the fundamental
transition decreases with increasing $N$, owing to the larger
density of states.\cite{QW} As a result, the phonon wavelength may
be too long to ever match anti-phase relation with the electron wave
function in the growth direction. In order to explore this issue, in
Fig.~\ref{Fig5} we represent the charge relaxation rate in a QD with
$\hbar \omega_0=5$ meV and $L_z=15$ nm with $N=2$ and $N=3$
electrons.
For $N=2$, two scattering minima show up at $B=0.6$ T and $B=2.4$ T.
These are the same values as found in the $N=1$ case,\cite{ClimentePRB}
which suggests that the $N=2$ scattering is well
described as an independent-particle event. 
However, the $N=3$ picture only shows one minimum at $B=0.55$ T
(the dip at $B \sim 4$ T is simply due to the crossing between the $(M=1,S=1/2)$
and $(M=2,S=1/2)$ energy levels).  
This is a clear manifestation of phonon energy reduction due to
Coulomb interactions. For a larger number of electrons (not shown),
when the density of states is still higher, the magnetic field
brings about frequent changes of energy levels with different
symmetry, so that tuning the emitted phonon energy is no longer
feasible. We then conclude that $B$-induced suppression of charge
relaxation rates holds for QDs with a small enough number of
electrons, but it rapidly loses efficiency as the number of
particles increases, owing to the increasing density of states.

\section{Coupled quantum dots}
\label{sec:cqd}

In this section we study charge relaxation rates corresponding to
$N$-electron isospin transitions in vertically CQDs as a function of
the interdot barrier thickness $L_b$. The two QDs are supposed to be
identical.\cite{asymmetric quantum dots} We focus on the transition
between the lowest symmetric ($G=0$) and antisymmetric ($G=1$)
solutions of the double quantum well with the spin quantum numbers
of the ground state, i.e., the fundamental isospin (interdot)
transition. This is indeed the fundamental transition of the system
when the tunneling energy is smaller than the lateral confinement
energy of the constituent QDs.

\begin{figure}[p]
\includegraphics[width=8cm,clip]{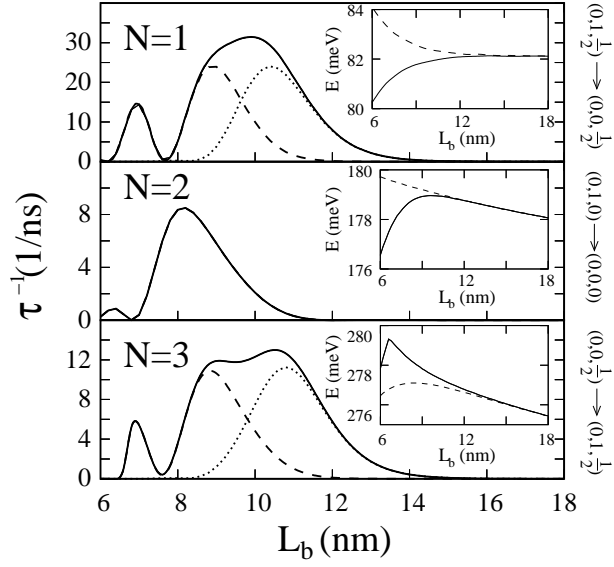}
\caption{Charge relaxation rate vs interdot barrier thickness $L_b$
in a CQD with $L_z=5$ nm, $\hbar \omega_0=5$ meV and $N=1,2,3$
electrons. Solid lines: total scattering rate. Dashed line: DP
contribution. Dotted lines: PZ contribution. Note the different
vertical scale in each panel. Next to the right axis, we indicate
the quantum numbers $(M,G,S)$ of the states involved in the
transition. For $N=2$, the PZ contribution cannot be distinguished
on this scale. Insets: lowest-lying symmetric (solid line) and
antisymmetric (dashed line) energy levels.}\label{Fig6}
\end{figure}

Figure \ref{Fig6} illustrates the results for a CQD structure with
$L_z=5$ nm, $\hbar \omega_0=5$ meV and $N=1,2,3$ electrons. Solid
lines indicate the total scattering rate, while dashed and dotted
lines represent the contribution coming separately from DP and PZ
interactions, respectively.
As for the intradot transition case, it is worth noting that the
$N=1$ calculation represents the independent-particle limit of the
$N=2$ and $N=3$ systems. Since the maximum scattering rate of the
$N=2$ ($N=3$) system is smaller than that of $N=1$ by a factor of
about 4 (2), we conclude that interdot transitions also benefit from a
correlation-induced reduction of the scattering rates. 
Such reduction can be interpreted in terms of configuration mixing
in analogous manner to the intradot case described in the previous section.

We note in Fig.~\ref{Fig6} that the shape of the scattering rate
curve is qualitatively similar for $N=1$ and $N=3$, with a
dominating and oscillating
 DP scattering intensity at small barrier thickness, and a PZ contribution which becomes dominant
as the barrier thickness exceeds $\sim 10$ nm.\cite{ClimentePRB} In
contrast, the shape significantly differs for $N=2$, where the PZ
contribution is missing. Actually, a close inspection reveals that
it has been suppressed by almost four orders of magnitude. This
striking result cannot be interpreted as a consequence of
correlations on the electron wave function, as in previous sections,
because the orbital part of the electron-phonon interaction matrix
element is identical for DP and PZ scattering mechanisms (see
Eq.~\ref{eq2}), and therefore the effect should be apparent also for
DP scattering. Therefore, the origin must be connected with the
effect of Coulomb interaction on the emitted phonon energy.

\begin{figure}[p]
\includegraphics[width=8cm,clip]{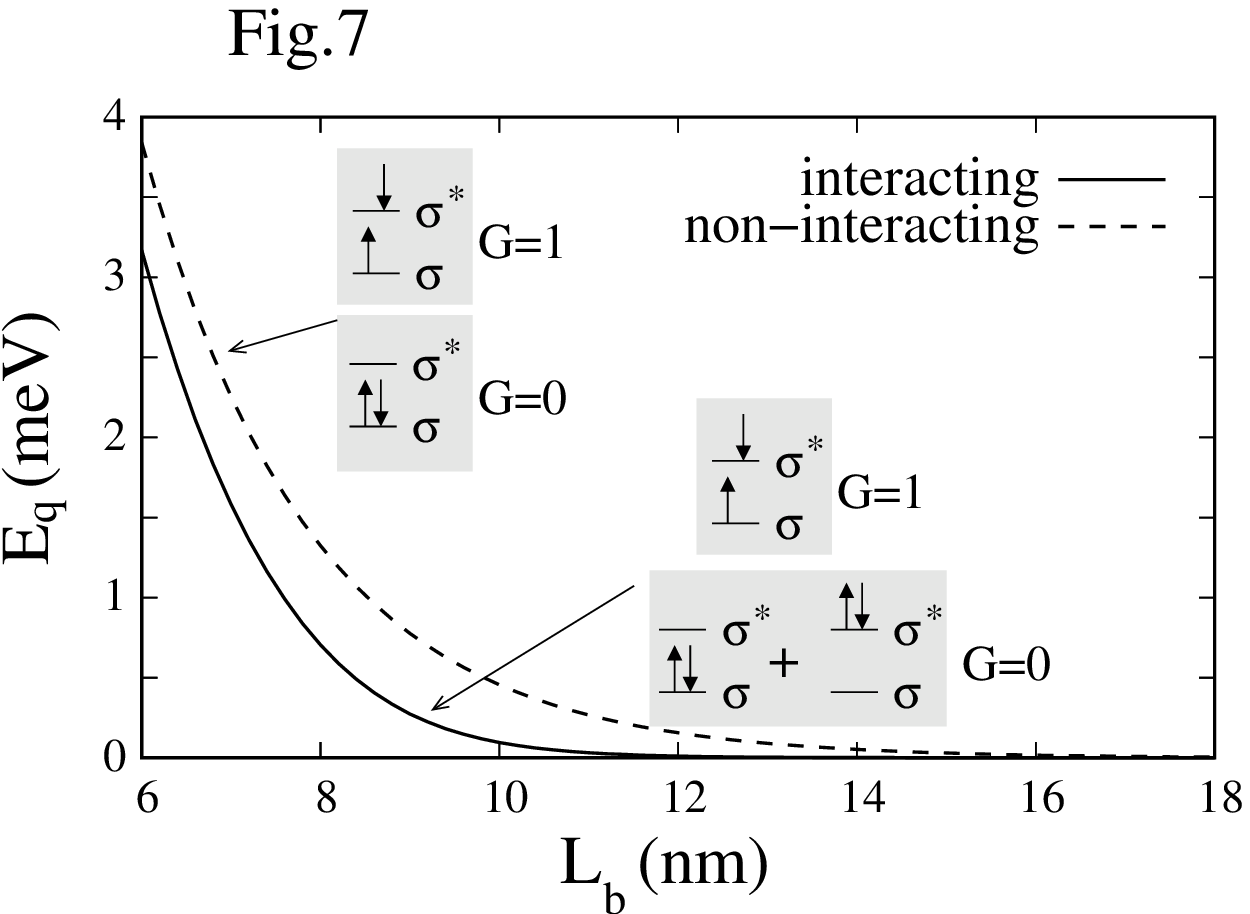}
\caption{Energy of the emitted phonon vs interdot barrier thickness
$L_b$ for $N=2$ interacting (solid line) and non-interacting (dashed
line) electrons in the CQD structure of Fig.~\ref{Fig6}. The
diagrams illustrate the dominant SE configurations with and without
Coulomb interaction.}\label{Fig7}
\end{figure}

To better understand this result, in Fig.~\ref{Fig7} 
we compare the $L_b$ dependence of the tunneling
energy, and hence that of the emitted phonon, for $N=2$ with and
without Coulomb interaction. Clearly, Coulomb interaction is
responsible for a significantly faster reduction of the tunneling
energy. Consequently, the values of $L_b$ which would lead to
maximum PZ scattering for non-interacting electrons (e.g.,
$L_b\approx 11$ nm), are now associated with very small phonon
energies. As a result, the phonon density of states is very small
and the electron-phonon coupling is strongly reduced.
This behavior is no longer found in the $N=3$ system. An
interpretation of this observation is provided in the diagrams of
Fig.~\ref{Fig7}, where we show the dominant SE configurations for
the interacting and non-interacting electrons. In the absence of
Coulomb interaction, the only configuration for the $G=0$ state has
 two electrons in the lowest SE symmetric orbital ($\sigma^2$).
Similarly, for the $G=1$ state the only significant configuration is that with
one electron in the lowest SE symmetric orbital and
another in the lowest antisymmetric orbital ($\sigma\,\sigma^*$).
However, when Coulomb interaction is included
for $G=0$ we mostly obtain a linear combination of $\sigma^2$ and
$(\sigma^*)^2$ configurations.
The thicker the barrier, the smaller the tunneling energy and the larger the
mixing between these two configurations.
In the limit where both configurations have equal weight, the $G=0$ state will
be degenerate with the $G=1$ one. In other words, Coulomb interaction
tends to make $G=0$ and $G=1$ solutions converge in energy with increasing $L_b$.
This is not possible for $N=3$ because the unpaired electron prevents a similar
mixing of configurations which use $\sigma$ and $\sigma^*$ orbitals while
preserving the total parity quantum number.
Therefore, the only significant electronic configuration for $G=0$ and $G=1$
are $\sigma^2 \sigma^*$ and $\sigma (\sigma^*)^2$, respectively.
The transition between these two configurations can be envisaged as the
relaxation of a single hole from $\sigma^*$ to $\sigma$ orbitals, which explains
the similar behavior of $N=3$ as compared to $N=1$.

\section{Summary}
\label{sec:summ}

We have investigated the effect of Coulomb interaction on the
charge relaxation rates of disk-shaped QD systems with few
electrons, due to incoherent coupling with acoustic phonons. We have
studied both intradot transitions in single QDs and interdot
(isospin) transitions in vertically CQDs. Coulomb interaction
affects the scattering rates in two ways: first, it changes the
emitted phonon energy and momentum, and second, it changes the
electron wave function through electronic correlations. Both effects
have a significant influence on the electron relaxation rate in a
way that generally leads to a reduction of the intradot and interdot
transition rates. This trend is gradually quenched by axial magnetic
fields. On the other hand, the increasing density of states with
higher number of particles reduces the energy of the emitted
phonons, which renders PZ interaction increasingly important, so
that  it may eventually become the dominant scattering mechanism
even at zero magnetic field. We have finally shown that the
suppression of electron scattering in weakly-confined QDs, which was
recently suggested for SE
systems,\cite{ClimentePRB,BertoniAPL,BertoniPE} also applies to
multi-electron QDs with a sufficiently small number of particles.

\begin{acknowledgments}
We acknowledge support from the Italian Ministry for University and
Scientific Research under FIRB RBIN04EY74, CINECA Calcolo parallelo 2006
and EU under the TMR network ``Exciting''.
\end{acknowledgments}

\clearpage

\pagebreak

\pagebreak

\pagebreak

\pagebreak

\pagebreak

\pagebreak

\begin{thebibliography}{30}
\bibitem{noseveN1}
The suppression of the relaxation rate at small confinement energies for $N=1$ is not seen in Figure
\ref{Fig1} because it occurs for characteristic frequencies $\hbar \omega_0 < 0.1$ meV,
but it is apparent e.g. in Figure \ref{Fig2}(c).
\bibitem{N2aN3}
The large shift from $N=2$ to $N=3$ is due to the occupation of the $p$-shell by the ground state electrons,
which greatly reduces the kinetic energy difference with respect to the excited states.
\bibitem{Ecte} It may be worth pointing out that the phonon energy of the plotted
 transitions in Figure \ref{Fig4} is identical, so the different rates are due to
orbital effects only.
\bibitem{QW} Note that the role of electronic correlations is secondary for this discussion
since we are concerned about the electron wave function in the vertical direction only,
which is the first solution of the quantum well for all low-lying QD states
due to the strong vertical confinement.
\bibitem{asymmetric quantum dots} The scattering rates of SE CQDs
with slightly anisotropic QDs have been studied in Ref. \onlinecite{ClimentePRB}.
\end{thebibliography}
\end{document}